%% file: main.tex
\documentclass[sigconf,nonacm]{acmart}
\usepackage{bbding}
\usepackage{color}
\usepackage{enumitem}
\usepackage{subcaption}
\AtBeginDocument{%
  \providecommand\BibTeX{{%
    \normalfont B\kern-0.5em{\scshape i\kern-0.25em b}\kern-0.8em\TeX}}}

\begin{document}

\title{A Dataset for the Validation of Truth Inference Algorithms Suitable for Online Deployment}

\author{Fei Wang}
\affiliation{%
	\institution{School of Computer Science and Technology, University of Science and Technology of China \& State Key Laboratory of Cognitive Intelligence}
	\city{Hefei}
	\country{China}
}
\email{wf314159@mail.ustc.edu.cn}
\author{Haoyu Liu}
\authornote{Corresponding author.}
\affiliation{%
	\institution{NetEase Inc.}
	\city{Hangzhou}
	\country{China}
}
\email{liuhaoyu03@corp.netease.com}
\author{Haoyang Bi}
\affiliation{%
	\institution{School of Computer Science and Technology, University of Science and Technology of China \& State Key Laboratory of Cognitive Intelligence}
	\city{Hefei}
	\country{China}
}
\email{bhy0521@mail.ustc.edu.cn}
\author{Xiangzhuang Shen}
\affiliation{%
	\institution{NetEase Inc.}
	\city{Hangzhou}
	\country{China}
}
\email{shenxiangzhuang@corp.netease.com}
\author{Renyu Zhu}
\affiliation{%
	\institution{NetEase Inc.}
	\city{Hangzhou}
	\country{China}
}
\email{zhurenyu@corp.netease.com}
\author{Runze Wu}
\authornotemark[1]
\affiliation{%
	\institution{NetEase Inc.}
	\city{Hangzhou}
	\country{China}
}
\email{wurunze1@corp.netease.com}
\author{Minmin Lin}
\affiliation{%
	\institution{NetEase Inc.}
	\city{Hangzhou}
	\country{China}
}
\email{linminmin01@corp.netease.com}
\author{Tangjie Lv}
\affiliation{%
	\institution{NetEase Inc.}
	\city{Hangzhou}
	\country{China}
}
\email{hzlvtangjie@corp.netease.com}
\author{Changjie Fan}
\affiliation{%
	\institution{NetEase Inc.}
	\city{Hangzhou}
	\country{China}
}
\email{fanchangjie@corp.netease.com}
\author{Qi Liu}
\affiliation{%
	\institution{School of Computer Science and Technology, University of Science and Technology of China \& State Key Laboratory of Cognitive Intelligence}
	\city{Hefei}
	\country{China}
}
\email{qiliuql@ustc.edu.cn}
\author{Zhenya Huang}
\affiliation{%
	\institution{School of Computer Science and Technology, University of Science and Technology of China \& State Key Laboratory of Cognitive Intelligence}
	\city{Hefei}
	\country{China}
}
\email{huangzhy@ustc.edu.cn}
\author{Enhong Chen}
\affiliation{%
	\department{}
	\institution{Anhui Province Key Laboratory of Big Data Analysis and Application, University of Science and Technology of China \& State Key Laboratory of Cognitive Intelligence}
	\city{Hefei}
	\country{China}
}
\email{cheneh@ustc.edu.cn}

\renewcommand{\shortauthors}{Fei Wang, et al.}

\begin{abstract}
	For the purpose of efficient and cost-effective large-scale data labeling, crowdsourcing is increasingly being utilized. To guarantee the quality of data labeling, multiple annotations need to be collected for each data sample, and truth inference algorithms have been developed to accurately infer the true labels. Despite previous studies having released public datasets to evaluate the efficacy of truth inference algorithms, these have typically focused on a single type of crowdsourcing task and neglected the temporal information associated with workers' annotation activities. These limitations significantly restrict the practical applicability of these algorithms, particularly in the context of long-term and online truth inference. In this paper, we introduce a substantial crowdsourcing annotation dataset collected from a real-world crowdsourcing platform. This dataset comprises approximately two thousand workers, one million tasks, and six million annotations. The data was gathered over a period of approximately six months from various types of tasks, and the timestamps of each annotation were preserved. We analyze the characteristics of the dataset from multiple perspectives and evaluate the effectiveness of several representative truth inference algorithms on this dataset. We anticipate that this dataset will stimulate future research on tracking workers' abilities over time in relation to different types of tasks, as well as enhancing online truth inference.
\end{abstract}





\maketitle

\input{introduction}

\input{relatedwork}

\input{data}

\input{experiments}

\input{conclusion}

\bibliographystyle{ACM-Reference-Format}
\bibliography{reference}

\appendix

\end{document}

%% file: introduction.tex
\section{Introduction}
Large-amount and high-quality data is the important and indispensable basis for machine learning or artificial intelligence algorithms, and labeling data samples is a frequently required step in data collection. In place of labeling through a few domain experts, crowdsourcing has been an effective solution that has the advantage of lower time and financial costs and higher flexibility \cite{zhang2016learning}. In data crowdsourcing, the annotators are likely to have less professional knowledge about the annotation tasks. Therefore, multiple annotations are collected for each data sample from different annotators, and then the most likely label (truth) is inferred from the annotations. This inferring process is the ``label aggregation'' stage on the crowdsourcing platform, and is also typically referred to as ``truth inference''. Truth inference is the crucial component of crowdsourcing, which ensures the quality of data labeling and thus has caught much attention.

\begin{figure}
    \centering
    \includegraphics[scale=0.4]{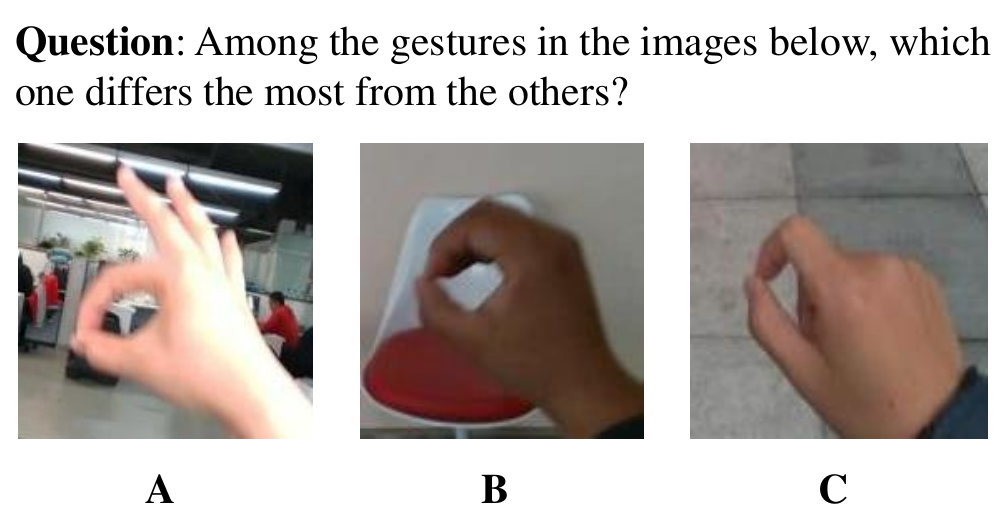}
    \caption{A task example in NetEaseCrowd.}
    \label{fig:task_example}
\end{figure}

The designing and evaluation of truth inference algorithms require corresponding data crowdsourcing datasets which include annotators, tasks, the annotated labels, and the true labels of the tasks. A task is typically a data sample to be labeled, and is displayed to the annotators as a question to be answered. Figure \ref{fig:task_example} demonstrates an example of task in our proposed dataset and how it is displayed to annotators. Table \ref{tab:dataset_comparison} lists some popular datasets used in the research and their basic statistics. Basically, each existing crowdsourcing dataset for truth inference is collected within a single type of task. Therefore, the data scale is relatively small, and the time spent on the labeling process is ignored. The annotators in such a dataset are most likely to be regarded as {\it temporary} annotators working for that set of crowdsourcing tasks, without recording the timestamp of their activities.

\begin{table*}[tb]
    \centering
    \caption{The comparison among existing datasets and NetEaseCrowd}
    \begin{tabular}{c|cccccccc}
        \toprule
        Dataset & \#Worker & \#Task & \#Groundtruth & \#Anno & Avg(\#Anno/worker) & Avg(\#Anno/task) & Timestamp & Task type \\
        \midrule
        Adult \cite{zheng2017truth} & 825 & 11,040 & 333 & 92,721 & 112.4 & 8.4 & \XSolidBrush & Single \\
        Birds \cite{zhang2016spectral} & 39 & 108 & 108 & 4,212 & 108.0 & 39.0 & \XSolidBrush & Single \\
        Dog \cite{yang2022light} & 109 & 807 & 807 & 8,070 & 74.0 & 10.0 & \XSolidBrush & Single \\
        CF \cite{li2019exploiting} & 461 & 300 & 300 & 1,720 & 3.7 & 5.7 & \XSolidBrush & Single \\
        CF\_amt \cite{li2019exploiting} & 110 & 300 & 300 & 6030 & 54.8 & 20.1 & \XSolidBrush & Single \\
        Emotion \cite{snow2008cheap} & 38 & 700 & 565 & 7,000 & 184.2 & 10.0 & \XSolidBrush & Single \\
        Smile \cite{whitehill2009whose} & 64 & 2,134 & 159 & 30,319 & 473.7 & 14.2 & \XSolidBrush & Single \\
        Face \cite{li2019exploiting} & 27 & 584 & 584 & 5,242 & 194.1 & 9.0 & \XSolidBrush & Single \\
        Fact \cite{yang2022light} & 57 & 42,624 & 576 & 216,725 & 3802.2 & 5.1 & \XSolidBrush & Single \\
        MS \cite{rodrigues2013learning} & 44 & 700 & 700 & 2,945 & 66.9 & 4.2 & \XSolidBrush & Single \\
        product \cite{wang2012crowder} & 176 & 8,315 & 8,315 & 24,945 & 141.7 & 3.0 & \XSolidBrush & Single \\
        RTE \cite{snow2008cheap} & 164 & 800 & 800 & 8,000 & 48.8 & 10.0 & \XSolidBrush & Single \\
        Sentiment \cite{venanzi2013bayesian} & 1,960 & 98,980 & 1,000 & 569,375 & 290.5 & 5.8 & \XSolidBrush & Single \\
        SP \cite{li2019exploiting} & 203 & 4,999 & 4,999 & 27,746 & 136.7 & 5.6 & \XSolidBrush & Single \\
        SP\_amt \cite{li2019exploiting} & 143 & 500 & 500 & 10,000 & 69.9 & 20.0 & \XSolidBrush & Single \\
        Trec \cite{lease2011overview} & 762 & 19,033 & 2,275 & 88,385 & 116.0 & 4.6 & \XSolidBrush & Single \\
        Tweet \cite{zheng2017truth} & 85 & 1,000 & 1,000 & 20,000 & 235.3 & 20.0 & \XSolidBrush & Single \\
        Web \cite{zhou2012learning} & 177 & 2,665 & 2,653 & 15,567 & 87.9 & 5.8 & \XSolidBrush & Single \\
        ZenCrowd\_us \cite{demartini2012zencrowd} & 74 & 2,040 & 2,040 & 12,190 & 164.7 & 6.0 & \XSolidBrush & Single \\
        ZenCrowd\_in \cite{demartini2012zencrowd} & 25 & 2,040 & 2,040 & 11,205 & 448.2 & 5.5 & \XSolidBrush & Single \\
        ZenCrowd\_all \cite{demartini2012zencrowd} & 78 & 2,040 & 2,040 & 21,855 & 280.2 & 10.7 & \XSolidBrush & Single \\
        \midrule
        \midrule
        NetEaseCrowd & 2,413 & 999,799 & 999,799 & 6,016,319 & 2,493.3 & 6.0 & \Checkmark & Multiple \\
        \bottomrule
    \end{tabular}
    \label{tab:dataset_comparison}
\end{table*}

However, with the increasing demand of data collection for large-scale and even more difficult tasks \cite{doroudi2016toward}, and the development of crowdsourcing platforms such as MTurk\footnote{\url{https://www.mturk.com/}}, CrowdFlower\footnote{\url{https://visit.figure-eight.com/People-Powered-Data-Enrichment_T}} and Scale AI\footnote{\url{https://scale.com/}}, the crowdsourcing tasks tend to gather to theses platforms, which employ {\it long-term} annotators (usually called workers) to conduct the labeling process. Consequently, a large amount of tasks are dispatched on the platforms to collect annotations and infer the truths. This raises some new challenges or opportunities of the truth inference algorithms, including:
\begin{itemize}
    \item The workers' annotation history of different tasks over time can be saved and used to better model the workers' abilities for inferring the truths in future tasks.
    \item There can be various task types on the crowdsourcing platform, which require different kinds of abilities to accomplish.
    \item There is the demand for online truth inference for applications such as task assignment \cite{hettiachchi2022survey} and active learning \cite{sayin2021review}. When the amount of online tasks is large, the efficiency of truth inference algorithms can be of concern.
\end{itemize}

To support the development of truth inference algorithms that settle these challenges, datasets are supposed to be large enough with the timestamp of each annotation recorded. Crowdsourcing data of different types of tasks will facilitate the exploration of how task type or different ability influences the truth inference process. Current public datasets that meet these requirements are scarce, which becomes the main obstacle of research along this line.

To this end, in this paper, we construct and release a new dataset, \textbf{NetEaseCrowd}, based on a mature data crowdsourcing platform of NetEase Inc. The dataset contains 2,413 workers, 999,799 tasks and 6 million annotations, where the annotations are collected in about 6 months. During the construction of the dataset, we deliberately chose more capabilities (6 level-3 capabilities), and each capability is associated with a sufficient number of relevant tasks. Worker IDs are anonymized to protect the user privacy. We analyze this dataset from several perspectives including workers, questions and annotations to show its differences and advantages compared with existing public datasets. Overall, NetEaseCrowd has the following advantages:
\begin{itemize}
    \item The task annotations are collected in a relatively long period, i.e., about 6 months, and the timestamp of each annotation is reserved.
    \item Multiple types of tasks are included and are arranged hierarchically. The tasks are published on the platforms in units of task sets, and each task set contains the same type of tasks relevant to the same capability.
    \item The dataset is large enough to test the efficiency of truth inference algorithms, which is important for online deployment.
\end{itemize}

Moreover, we test popular truth inference models on NetEaseCrowd with extensive experiments, which manifests the influence of the task capability and workers' temporal changes of features on truth inference. Then, we analyze the inference efficiency of the models. The experimental results reveal that supervised models are promising to inference online considering their efficiency, although further study is required to improve their accuracy when inferring without model retraining. Finally, we discuss potential research directions based on NetEaseCrowd, such as supervised truth inference algorithms, fine-grained worker ability estimation and tracking.

This dataset is ready to use with documentation at \url{https://github.com/fuxiAIlab/NetEaseCrowd-Dataset}, and is licensed under CC BY-SA 4.0\footnote{\url{https://creativecommons.org/licenses/by-sa/4.0/}}

%% file: relatedwork.tex
\section{Related Work}
There has been a considerable amount of public datasets for truth inference. Most of these datasets collect annotations for decision-making tasks or single-choice tasks. Popular datasets involve different types of tasks, such as target recognition (Dog \cite{zhou2012learning}, Bird \cite{zhang2016spectral}, etc.), natural language analysis (RTE \cite{snow2008cheap}, Trec \cite{lease2011overview}, Sentiment \cite{venanzi2013bayesian}, etc.), web content judgment (Web \cite{zhou2012learning}, Adult \cite{zheng2017truth}, etc.), and entity linking (ZenCrowd\_us, ZenCrowd\_in, ZenCrowd\_all \cite{demartini2012zencrowd}, etc.). There are also several datasets containing numerical tasks, such as Emotion \cite{snow2008cheap}, Population and Bio \cite{zhao2012probabilistic}. The annotations of these datasets are usually collected by publishing the tasks on a crowdsourcing platform such as the Amazon Mechanical Turk. 

Mostly, existing datasets only contains basic information for truth inference, i.e., worker IDs, task IDs, task truths and task annotations, while some of them also provide the task content, e.g., the texts of images to be annotated\footnote{\url{https://zhydhkcws.github.io/crowd_truth_inference/datasets.zip}}. Only a few datasets provide extra information relevant to the annotations. For example, ZenCrowd\_us, ZenCrowd\_in and ZenCrowd\_all provide the time spent on each annotation. % Need to be confirmed. The original url of the dataset is expired. This is inferred from the processing code https://github.com/orchidproject/active-crowd-toolkit/blob/master/CrowdsourcingModels/Datum.cs#L49
Table \ref{tab:dataset_comparison} lists some existing public datasets used in truth inference research papers. An important characteristic of these datasets is that they are collected based on a specific set of tasks, which means that the information retained is limited to that task set. As a result, the behaviors of the workers on historical tasks are unavailable, and the time information of the annotations is ignored, leading to relatively small and limited datasets. By contrast, the NetEaseCrowd we propose in this paper is a large crowdsourcing dataset containing long-term workers' annotations on multiple types of tasks over about six months. The timestamps of the annotations are reserved, which makes it possible to model the temporal features of workers.

%% file: data.tex
\begin{figure*}[htb]
	\centering
	\includegraphics[scale=0.5]{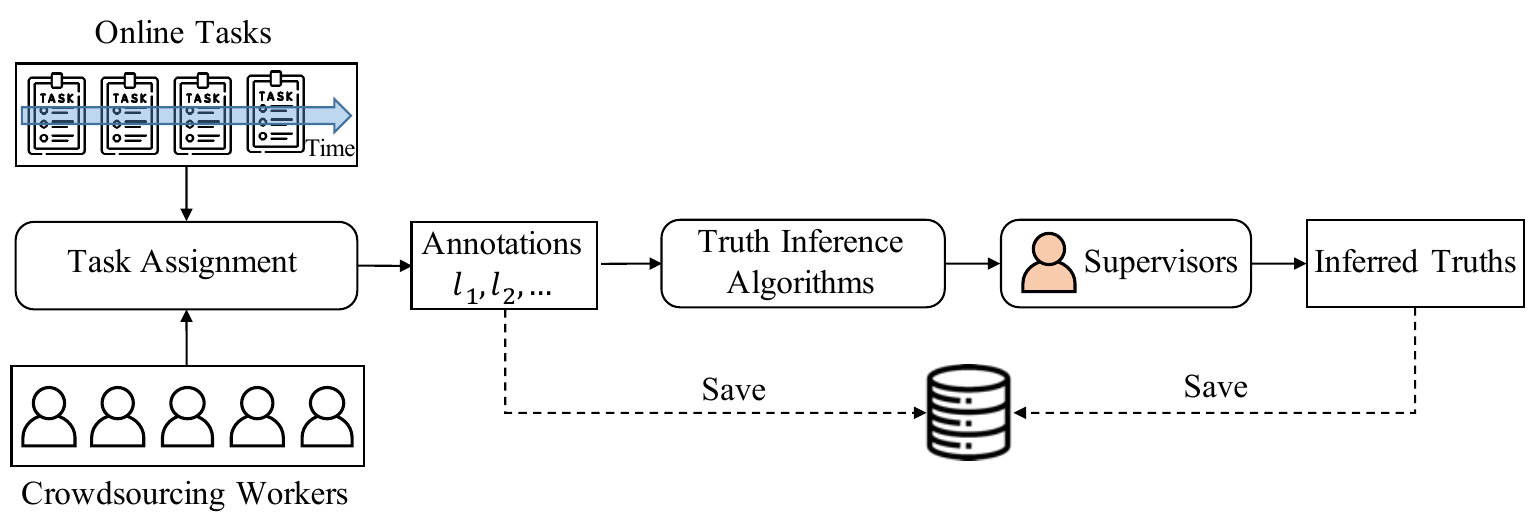}
	\caption{The procedure of data labeling on the crowdsourcing platform.}
	\label{fig:platform_process}
\end{figure*}
\section{Data Description}
The dataset we propose is collected from the data crowdsourcing platform from NetEase Inc. Normally, the data labeling process is illustrated in Figure \ref{fig:platform_process}. There are workers continuously working on the platform as annotators, and new tasks (i.e., data samples to be labeled) are constantly posted online. After task assignment, each task is labeled by multiple workers. Then, truths of the tasks are inferred from the annotated labels through both truth inference algorithms and supervisors with expert knowledge. The data, including tasks' annotated labels and their inferred truths, are saved.

Different from the existing public datasets which merely include crowdsourcing data collected for a single type of task, various types of tasks are published on the crowdsourcing platform over time. Therefore, the dataset \textbf{NetEaseCrowd} we construct based on the saved data better reflects the demands and characteristics of truth inference on a crowdsourcing platform. The basic statistics are presented in Table \ref{tab:dataset_comparison}. Specifically, NetEaseCrowd contains multiple types of tasks requiring different capabilities. Workers' performances over a longer period of time, i.e., about six months, are reserved together with their timestamps. Therefore, the sequential characteristics of online truth inference can be explored, such as the changes of worker abilities and task types. Moreover, the size of NetEaseCrowd is large enough for researchers to evaluate the efficiency of their algorithms. The details of our data collection and process are presented in the following subsections.
\begin{figure}[!tb]
    \centering
    
    \begin{subfigure}{\linewidth}
        \includegraphics[width=\linewidth]{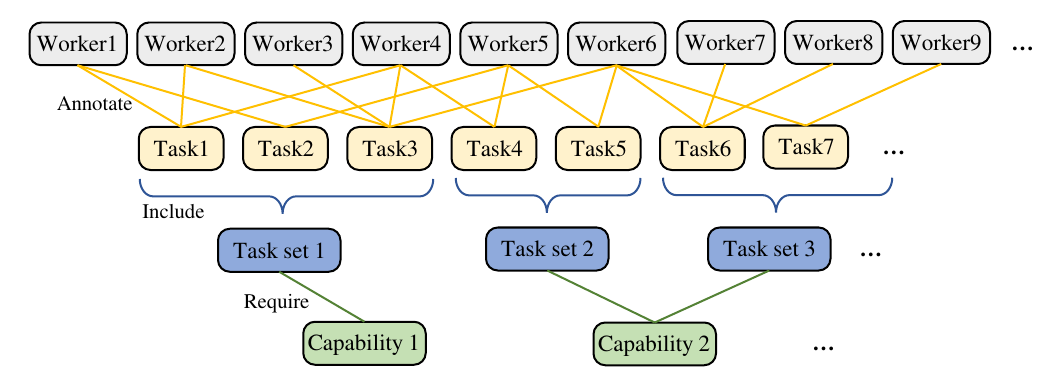}
        \caption{The relationship among capability, task set, task and worker.}
        \label{fig:cap_taskset_task_worker}
    \end{subfigure}
    
    \begin{subfigure}{\linewidth}
        \includegraphics[width=0.9\linewidth]{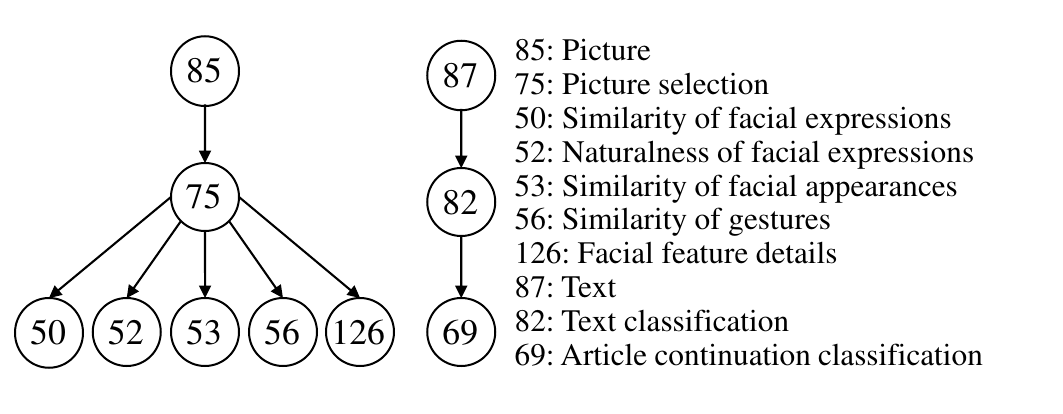}
        \caption{The structure of capabilities.}
        \label{fig:capability}
    \end{subfigure}
    
    \caption{Data structure of NetEaseCrowd.}
    \label{fig:data_structure}
\end{figure}

\subsection{Data Construction}
In our crowdsourcing platform, tasks are released in sets, where each task set contains the same type of tasks requiring the same capability of workers (e.g., appearance similarity judgment). The annotations of one task set are collected in a short time, such as one day. Figure \ref{fig:cap_taskset_task_worker} is a sketch of the relationships, and the capabilities in our platform are arranged to have three levels as shown in Figure \ref{fig:capability}. The construction process of NetEaseCrowd is as follows:
\begin{enumerate}[label={Step \theenumi.}]
    \item \textbf{Task Set Collection.} For each 3rd-level capability, we search its corresponding task sets. Each task set is identified with a unique ``TasksetId''. Only task sets that have been finished and passed the manual quality inspection are reserved.
    \item \textbf{Task Set Filtering.} For each 3rd-level capability, we reserve at most 100 relevant task sets, whose tasks are three-choice questions and are completed before 09/01/2023. When there are more than 100 task sets for a 3rd-level capability in the previous step, we reserve the task sets completed last.
    \item \textbf{Annotation Collection.} For each task set obtained in the previous step, we collect the annotated labels and inferred truths from our database. It should be noted that, the number of annotations collected for each tasks varies according multiple rules to ensure the quality of task truths. We randomly select no more than 10 annotations for each tasks to encourage studies that can achieve better truth inference accuracy with fewer annotations, which is meaningful for both academic research and business application. 
    \item \textbf{Anonymization.} Since the dataset is collected from a commercial crowdsourcing platform, we need to anonymize NetEaseCrowd to avoid possible private information leakage. The worker IDs in NetEaseCrowd have been anonymized. As the content of the tasks may involve the rights of data requesters, we do not provide them in NetEaseCrowd.
\end{enumerate}

\subsection{Data Analysis}
The basic statistics of NetEaseCrowd have been listed in Table \ref{tab:dataset_comparison}. We can observe from the table that compared to existing public datasets, NetEaseCrowd is a much larger dataset, including more workers, tasks and annotated labels. On average, each worker annotated 2,493.3 tasks, which is significantly more than most of the existing public datasets. The average amount of annotations per task is 6.0, reflecting the real-world demand by crowdsourcing platforms to obtain the task truths with fewer annotations. Moreover, NetEaseCrowd reserves the timestamp of each annotation and contains multiple types of tasks. 
\begin{figure}[!tb]
    \centering
    
    \begin{subfigure}{\linewidth}
        \includegraphics[width=0.95\linewidth]{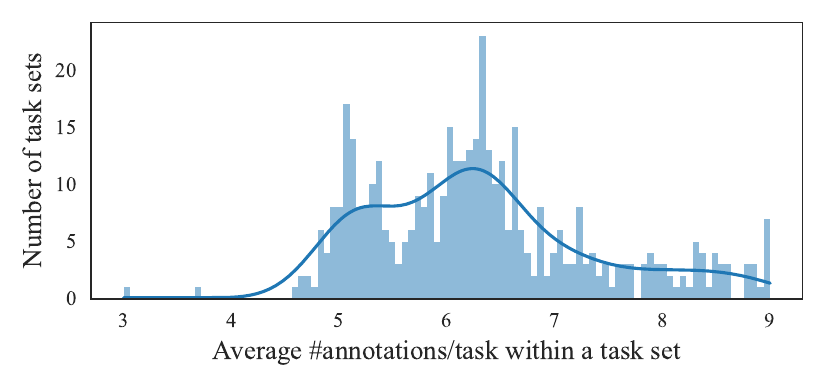}
        \caption{The distribution of \#annotation/task.}
        \label{fig:task_anno_dist}
    \end{subfigure}
    
    \begin{subfigure}{\linewidth}
        \includegraphics[width=0.95\linewidth]{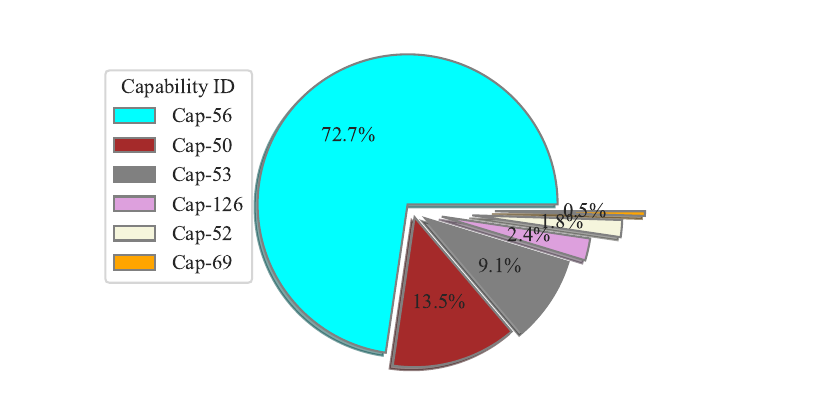}
        \caption{The distribution of tasks relevant to each capability.}
        \label{fig:cap_task_dist}
    \end{subfigure}
    
    \caption{Basic statistical properties of tasks.}
    \label{fig:task_property}
\end{figure}

\subsubsection{Task properties}
As presented in Table \ref{tab:dataset_comparison}, NetEaseCrowd contains 999,799 tasks, and the average number of annotations collected for each task is 6.0. In this part, we will provide some statistics at the level of task set and capability. First, we calculate the number of annotations provided to each task and average these numbers within each task set. The distribution of the averaged \#annotations/task is presented in Figure \ref{fig:task_anno_dist}. As described in data construction, when a task receives more than 10 annotated labels in the original data, we randomly keep 10 annotations. From Figure \ref{fig:task_anno_dist} we can observe that most task sets ask for 4$\sim$9 workers to annotate for a task. This might not be a large number annotation if compared to some research datasets (e.g., Birds \cite{zhang2016spectral} and CF\_amt \cite{li2019exploiting}); however, this truly reflects the annotation volume accepted by a real-world data crowdsourcing platform and raises requirements for accurate truth inference algorithms under low annotation volume. Then, as there are multiple types of tasks requiring different worker capabilities, we illustrate the distribution of tasks relevant to each capability. As shown in Figure \ref{fig:cap_task_dist}, the majority of tasks in NetEaseCrowd are related to ``Similarity of gestures'' (ID=56), followed by ``Similarity of facial expressions'' and ``Similarity of facial appearances'' (ID=50, 53). When estimating workers' abilities in truth inference algorithms, discriminating the abilities according to the relevant capabilities should improve the accuracy of inference. We will conduct experiments and discuss the influence of task set and capability on the accuracy of truth inference in \ref{sec:exp_effective}.

\subsubsection{Annotation properties}
\begin{figure}[!tb]
    \centering
    
    \begin{subfigure}{\linewidth}
        \includegraphics[width=0.95\linewidth]{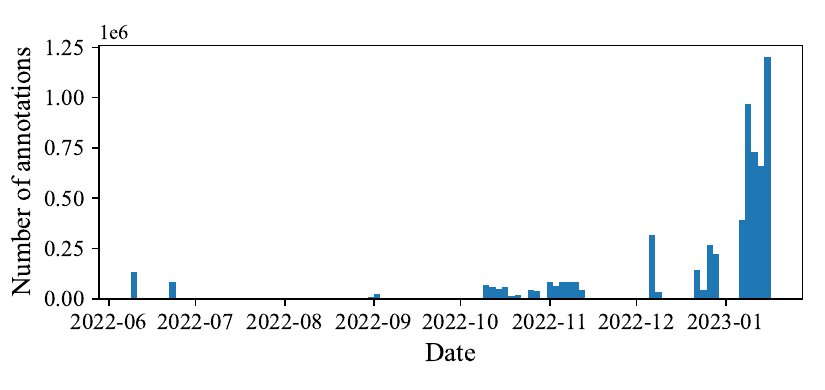}
        \caption{The number of annotations over time.}
        \label{fig:ano_time_dist}
    \end{subfigure}
    
    \begin{subfigure}{\linewidth}
        \includegraphics[width=0.95\linewidth]{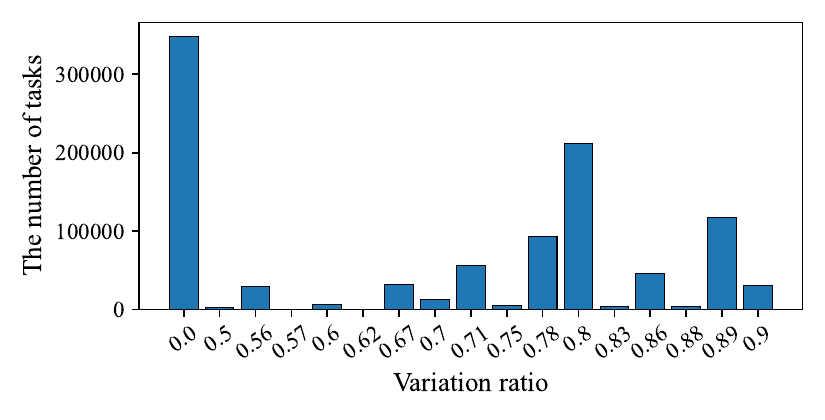}
        \caption{The variation ratios of tasks.}
        \label{fig:variation_ratio}
    \end{subfigure}
    
    \caption{Basic statistical properties of annotations.}
    \label{fig:worker_property}
\end{figure}

In Figure \ref{fig:ano_time_dist} we present the time distribution of workers' annotations. The annotations are distributed over the period of time from June 2022 to February 2023, and most annotations are collected during the period from October 2022 to February 2023. Then, to demonstrate the dispersion of task annotations, we calculate the variation ratio \cite{freeman1965elementary} of each task and present its distribution in Figure \ref{fig:variation_ratio}. The variation ratio of a task is calculated as:
\begin{equation}
    v:= 1 - \frac{f_m}{N},
\end{equation}
where $f_m$ is the frequency of the mode (the most voted choice), and $N$ is the number of annotations received by this task. The larger the variation ratio is, the more dispersed the annotations are. From Figure \ref{fig:variation_ratio} we observe that there are quite a lot of tasks whose variation ratios are larger than 0.67. This implies that the tasks in NetEaseCrowd possess a certain level of difficulty, leading to inconsistencies in the opinions of workers.

\iffalse
\begin{figure}
    \centering
    \includegraphics[scale=0.5]{anno_variation.pdf}
    \caption{The variation of annotations provided to questions. (forgot to set ddof=0)}
    \label{fig:variation}
\end{figure}
\fi

\subsubsection{Worker properties} \label{sec:worker_property}
In this part, we present some basic overview of workers' annotation behavior in NetEaseCrowd, and then provide some statistical analysis about the variation of worker ability over time and capability. 

\begin{figure}[!tb]
    \centering
    
    \begin{subfigure}{\linewidth}
        \includegraphics[width=0.95\linewidth]{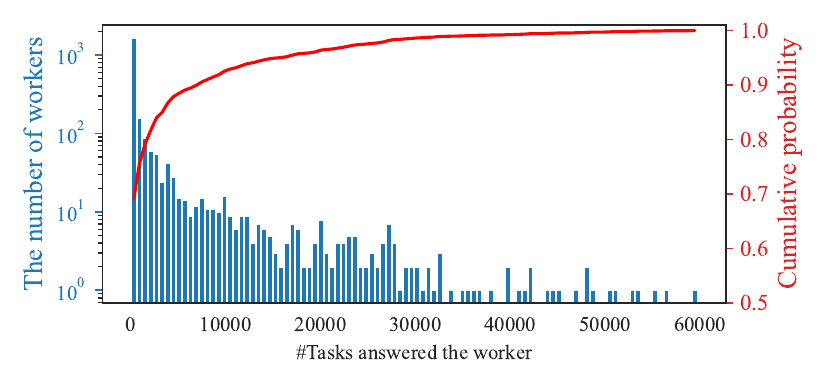}
        \caption{The distribution of \#annotation/worker.}
        \label{fig:worker_anno_dist}
    \end{subfigure}
    
    \begin{subfigure}{\linewidth}
        \includegraphics[width=0.95\linewidth]{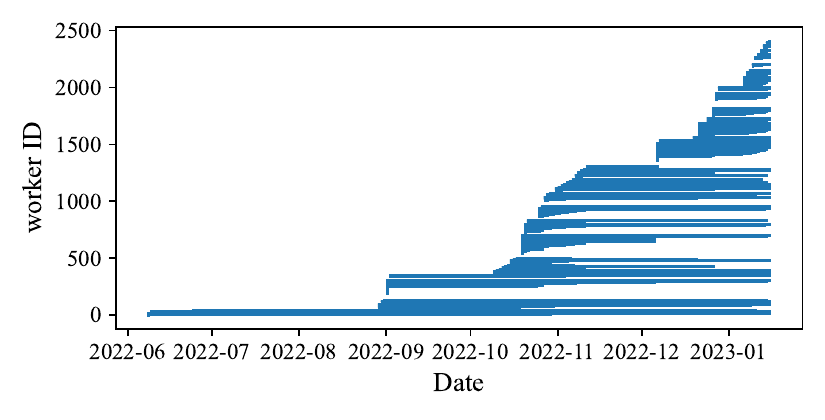}
        \caption{Workers' active time periods.}
        \label{fig:worker_date}
    \end{subfigure}
    
    \caption{Basic statistical properties of workers.}
    \label{fig:worker_property}
\end{figure}

First, we demonstrate in Figure \ref{fig:worker_anno_dist} the distribution of the numbers of annotations provided by every worker. Most workers in NetEaseCrowd participated in large amounts of tasks. Nevertheless, there are still considerable workers provided relatively small amounts of annotations, and this indicates that research about the cold start problem of worker ability evaluation is desirable. We draw the active time periods of each worker in Figure \ref{fig:worker_date}. From the figure we can have a clear overview that there are increasingly new workers on the crowdsourcing platform. The cold start problem in Figure \ref{fig:worker_anno_dist} results from both workers registered late and workers registered early but inactive.

\begin{figure}[tb]
    \centering
    \includegraphics[width=0.4\textwidth]{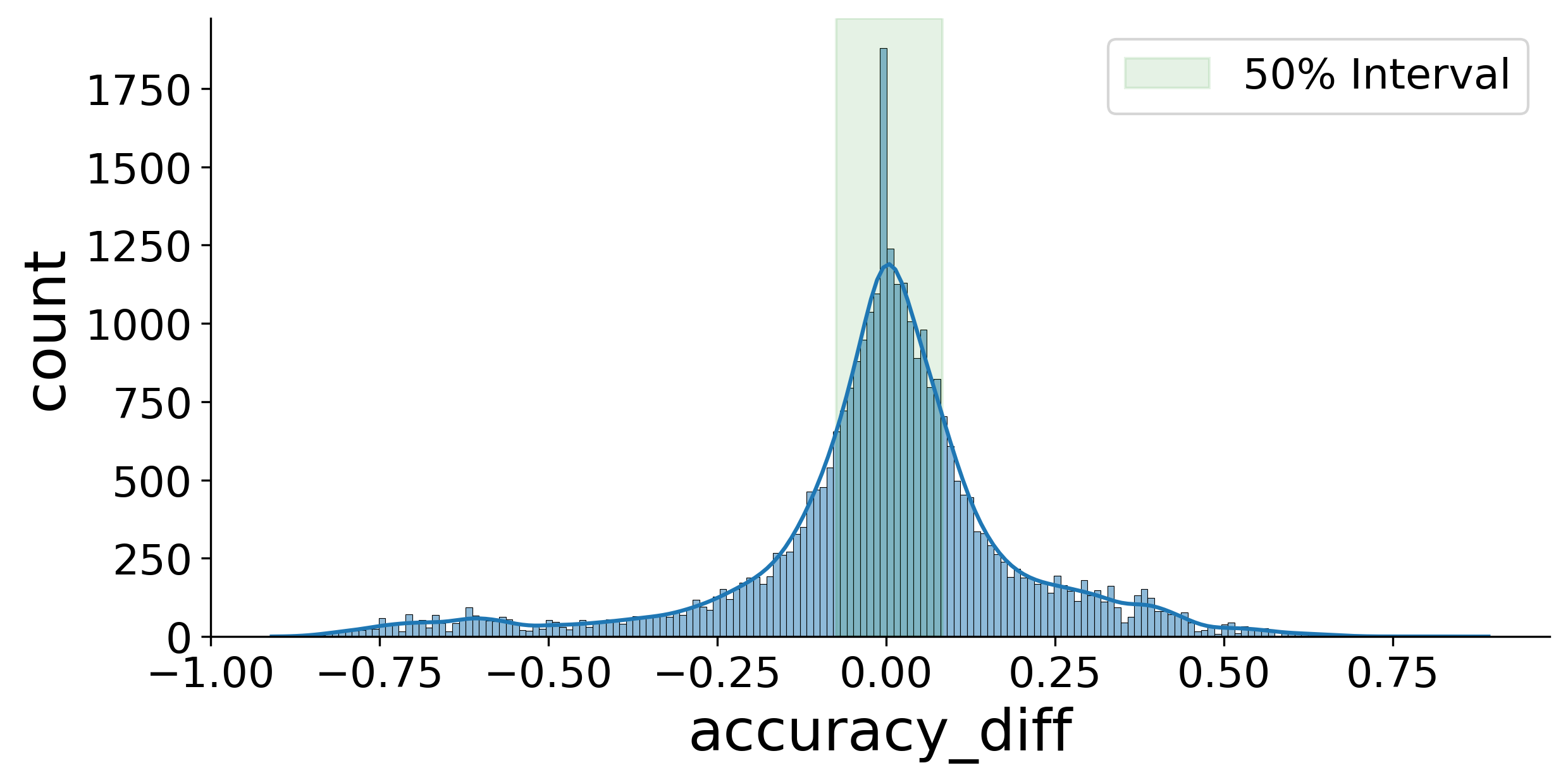}
    \caption{Accuracy differences distribution for all workers.}
    \label{fig: analysis-capwise-accuracy-diff}
\end{figure}

\begin{figure}[!htb]
    \centering
    
    \begin{subfigure}{0.9\linewidth}
        \includegraphics[width=0.95\linewidth]{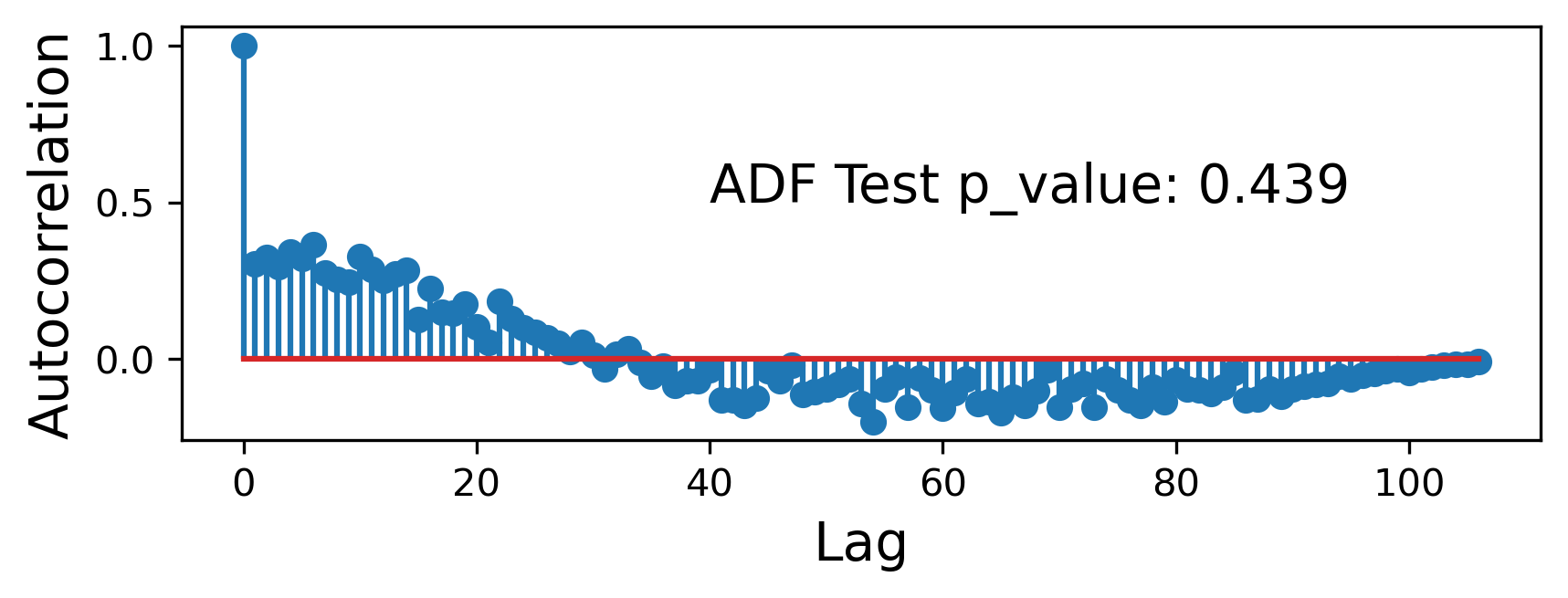}
        \caption{Autocorrelation analysis for worker 7477.}
        % \label{fig:redundancy_1}
    \end{subfigure}
    
    \begin{subfigure}{0.9\linewidth}
        \includegraphics[width=0.95\linewidth]{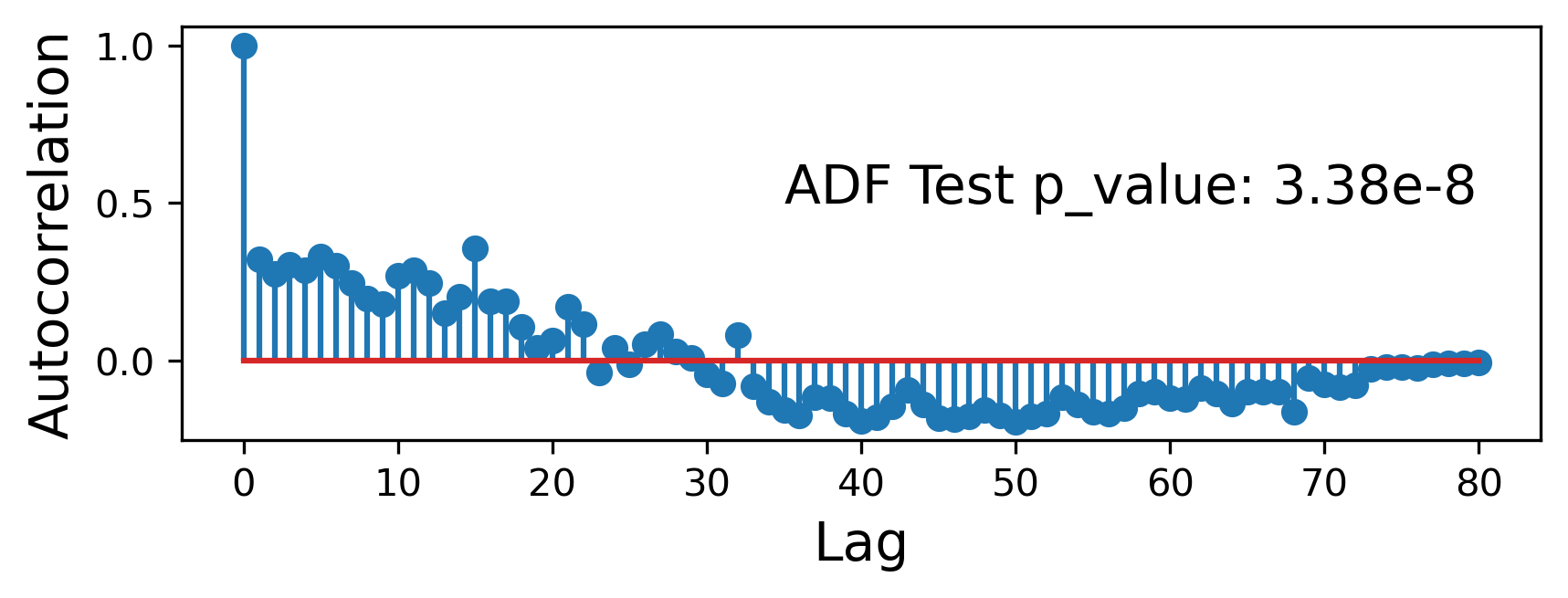}
        \caption{Autocorrelation analysis for worker 13600.}
        % \label{fig:redundancy_1}
    \end{subfigure}
    
    \caption{ACF annd ADF Test results.}
    \label{fig: autocorrelation_analysis}
\end{figure}

Then, we provide statistical insights into workers' label abilities over time. We calculate each worker's overall accuracy on all the answered tasks, and subtract it by the worker's accuracy on each answered task set. The distribution of all accuracy differences is illustrated in Figure~\ref{fig: analysis-capwise-accuracy-diff}. As depicted in the figure, accuracy differences can exceed 0.5 in some task sets. This highlights the inadequacy of using an overall compromised worker abilities to infer task labels in different time periods. Additionally, we selecte two workers with over 50,000 annotation records to illustrate the temporal features of worker accuracy. The Autocorrelation Function (ACF) is employed to quantify the correlation between data points at different time lags. We calculate the accuracy of workers in each task set and use task set-wise lag alongside time dimension to compute ACF. Each lag represents one next task set in chronological order. The results are presented in Figure~\ref{fig: autocorrelation_analysis},
which shows that when the lag is small, the ACF values are over 0.3, which represents a moderate positive correlation, suggesting the presence of evolving temporal features in worker abilities. We further provide the p-value of the Augmented Dickey-Fuller Test for these two workers, with the results displayed in Figure~\ref{fig: autocorrelation_analysis}. A p-value greater than 0.05 indicates nonstationary temporal features, while a p-value less than or equal to 0.05 suggests stationary temporal features. The results suggest that workers in NetEaseCrowd have diverse temporal features in their abilities.

\begin{figure}[!htb]
    \centering
    
    \begin{subfigure}{0.45\linewidth}
        \includegraphics[width=0.95\linewidth, height=5cm]{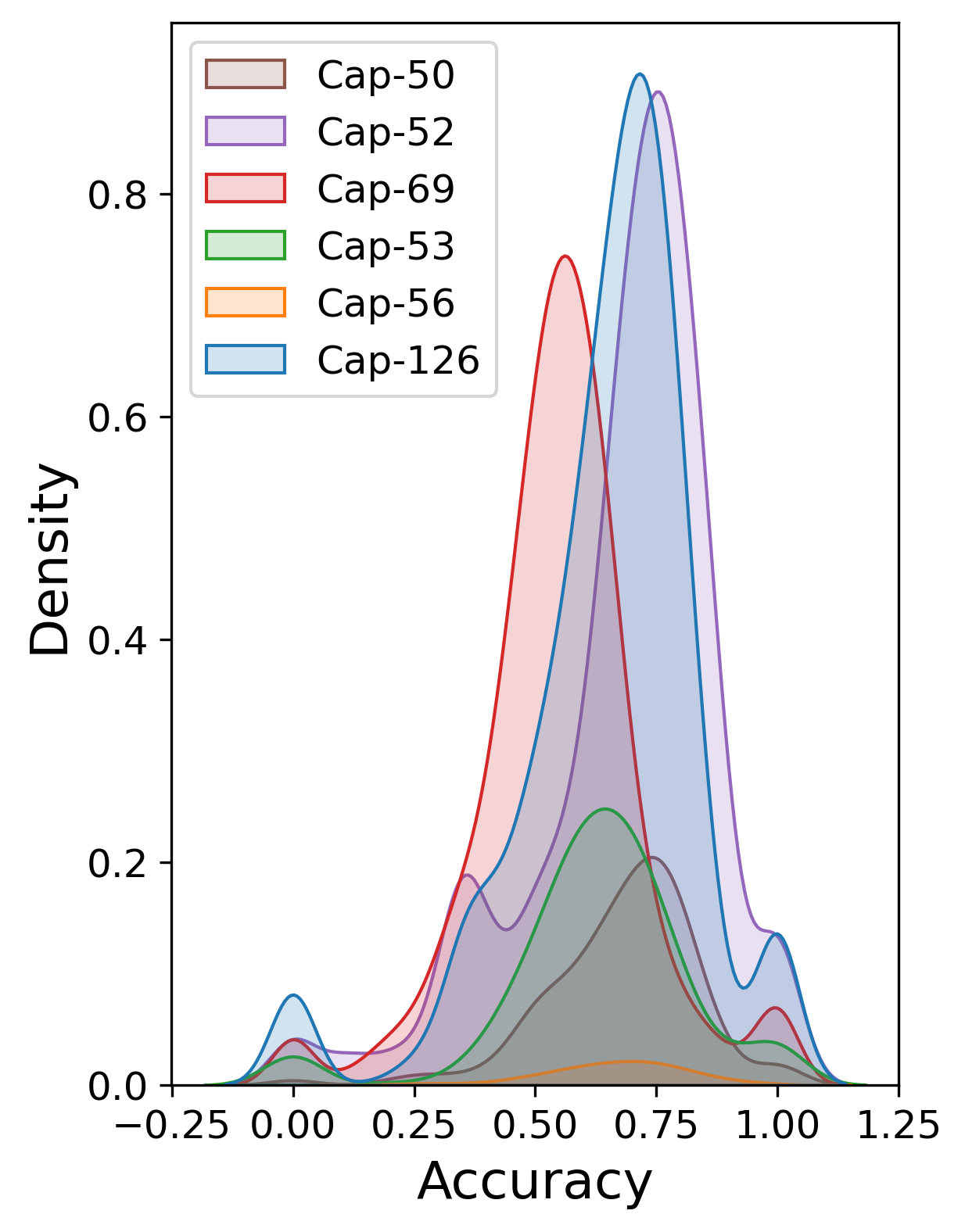}
        \caption{Capability-wise worker accuracy kernal density estimate results.}
        \label{fig:cap-wise-distribution}
    \end{subfigure}
    \begin{subfigure}{0.45\linewidth}
        \includegraphics[width=0.95\linewidth, height=5cm]{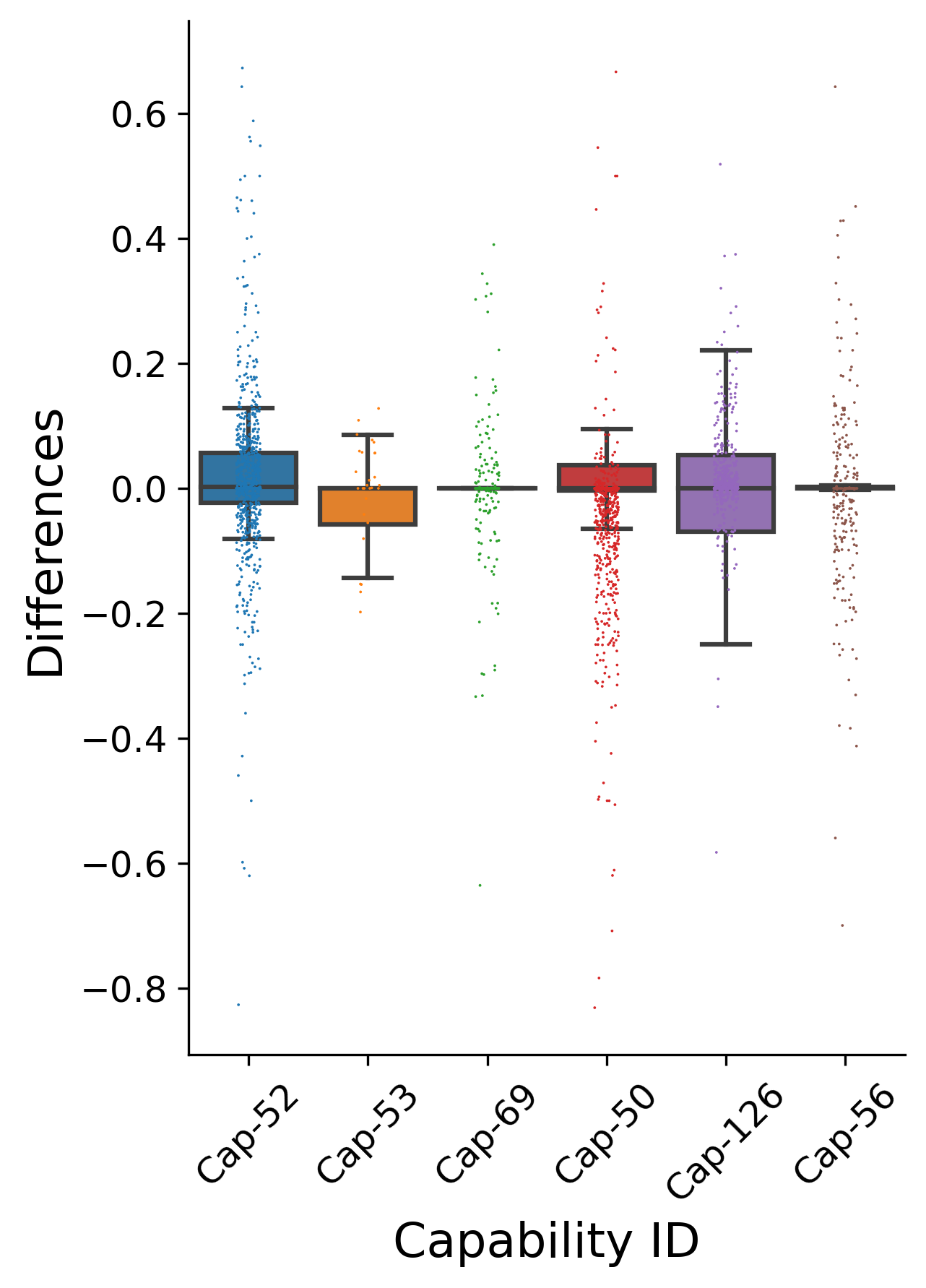}
        \caption{Distributions of differences between overall accuracy and capability-wise accuracy.}
        \label{fig:cap-wise-diff}
    \end{subfigure}

    \caption{Workers' accuracy within difference capabilities.}
    \label{fig: cap-wise-accuracy-features}
\end{figure}

Finally, among different capabilities, workers will exhibit distinct label abilities. We visualize the accuracy distribution of all workers on various capabilities in Figure~\ref{fig:cap-wise-distribution}, which illustrates that workers' accuracy forms diverse distributions within different capabilities. To compare the worker abilities within a single capability to those in the overall dataset, we present Figure~\ref{fig:cap-wise-diff}, showcasing the subtraction between workers' accuracy in the overall dataset and each sub-dataset with the same capability task sets. It indicates that, although within several capabilities, the worker accuracy has relatively close differences with that in the overall dataset, there are still some capabilities showing large differences. These results suggest that capability-wise information is also a critical feature for a fine-grained worker ability model and better label aggregation.

%% file: experiments.tex
\section{Experiments And Analysis}
NetEaseCrowd dataset can be used for investigations into new truth inference methods, particularly for exploring temporal features and the graph structure among workers, tasks, and various capabilities. In this section, we evaluate previous truth inference methods in NetEaseCrowd dataset and its variations to illustrate the dataset's effectiveness. 

\subsection{Experimental Setup}

\subsubsection{Truth inference methods}
We provide a comprehensive exploration over previous methods, where 11 benchmark methods are employed, including majority voting methods MV, WAWA~\cite{crowdkit2023} and ZeroBasedSkill~\cite{crowdkit2023}, probabilistic methods DS~\cite{dawid1979maximum}, GLAD~\cite{whitehill2009whose}, ZC~\cite{demartini2012zencrowd}, MACE~\cite{hovy2013learning}, EBCC~\cite{li2019exploiting}, neural network based methods LAA~\cite{li2017aggregating}, BiLA~\cite{hong2021online}, and TiReMGE~\cite{wu2023crowdsourcing}. 

\begin{table*}[!ht]
	\centering
	\caption{Performance on \textit{set-wise}, \textit{cap-wise}, and \textit{overall} verisions of datasets.}
	\begin{tabular}{cccccccc}
		\toprule
		& \multicolumn{3}{c}{Accuracy} &\multicolumn{3}{c}{F1-score} \\ \hline
		Variation & Set-wise & Cap-wise & Overall & Set-wise & Cap-wise & Overall \\\hline
		MV & 0.92695 & 0.92695 & 0.92695 & 0.92692 & 0.92692 & 0.92692 \\
		DS & 0.95180 & 0.94548 & 0.94820 & 0.95178 & 0.94552 & 0.94817 \\ 
		MACE & 0.95993 & 0.95287 & 0.94962 & 0.95991 & 0.95283 & 0.94957 \\
		Wawa & 0.94820 & 0.94555 & 0.94451 & 0.94814 & 0.94549 & 0.94445 \\
		ZeroBasedSkill & 0.94903 & 0.94703 & 0.94898 & 0.94898 & 0.94697 & 0.94585 \\ 
		GLAD & 0.95619 & 0.95365 & 0.95064 & 0.95615 & 0.95359 & 0.95058 \\ 
		EBCC & 0.93710 & 0.94318 & 0.91071 & 0.93707 & 0.94316 & 0.90996 \\ 
		ZC & 0.96705 & 0.95443 & 0.95305 & 0.96702 & 0.95439 & 0.95301 \\ 
		TiReMGE & 0.92753 & 0.92759 & 0.92713 & 0.92746 & 0.92754 & 0.92706 \\ 
		LAA & 0.93754 & 0.94061 & 0.94173 & 0.93756 & 0.94068 & 0.94169 \\
		BiLA & 0.89209 & 0.87709 & 0.88036 & 0.89265 & 0.87606 & 0.87896 \\\hline
		\textbf{Cohen's d with overall} & \textbf{0.35310} & \textbf{0.16235} & \textbf{0.0} & \textbf{0.36394} & \textbf{0.16567} & \textbf{0.0} \\ \bottomrule
	\end{tabular}
	\label{tab: overall_performance}
	\vspace{-0.1in}
\end{table*}

\subsubsection{Implementation and Metrics}
Without being specifically mentioned, these methods are all used in an unsupervised manner, i.e., training and inference on the same part of the dataset. 
All methods are implemented based on the hyperparameters outlined in their original papers. For evaluation metrics, we use accuracy and f1-score to measure the aggregation performance. All experiments are conducted on a server with Intel(R) Xeon(R) E5-2680 v4 @ 2.40GHz CPU and one NVIDIA GeForce GTX 1080 GPU.

\subsubsection{Dataset Setup}
NetEaseCrowd is formalized into four variations to explore its different features:
\begin{itemize}
    \item \textit{Overall:} Methods are applied to the entire NetEaseCrowd dataset. The inputs are all annotations and the outputs are the inferred labels.
    \item \textit{Set-wise:} The methods are applied to each sub-dataset, each of which contains tasks with the same task set IDs. Labels are inferred independently within each sub-dataset. The inferred labels for the entire NetEaseCrowd dataset consist of the concatenation of the inferred labels from every sub-dataset.
    \item \textit{Cap-wise:} The methods are applied to each sub-dataset, which contains tasks with the same capability IDs. The inferred labels are obtained in the same manner as in the Set-wise setting.
    \item \textit{Supervised:} NetEaseCrowd is divided into a training set and a testing set, alongside timestamps, comprising 75\% and 25\% of the total dataset, respectively. Performance is evaluated solely on the testing dataset. Unsupervised methods are trained and tested using only the testing dataset, while supervised methods are initially trained on the training dataset and subsequently tested on the testing dataset.
\end{itemize}
Note that overall, set-wise, and cap-wise variations are all designed for unsupervised inference, which is the most widely adopted inference type among all previous truth inference methods.

\subsection{Experimental Analysis}
In this subsection, we aim to answer four questions:
\begin{itemize}
    \item \textbf{Q1}: Does NetEaseCrowd contain inherent relations among workers, tasks, and annotations that models can learn, thereby improving aggregation results beyond trivial MV?

    \item \textbf{Q2}: Is it possible to use the provided temporal information for better worker modeling and to improve the inference performance?
    
    \item \textbf{Q3}: Are the provided capability IDs relevant to the workers' abilities and the inference performance?

    \item \textbf{Q4}: Is NetEaseCrowd an appropriate dataset for providing a supervised truth inference scenario?
\end{itemize}

Table~\ref{tab: overall_performance} displays the results of baseline methods applied to the dataset with various variations. Each method is employed under each dataset variation, and both accuracy and F1-score are reported. Additionally, Cohen's d value is calculated, with set-wise and cap-wise variations utilized as treatment groups, and the overall variation as the control group. A positive Cohen's d value indicates that the treatment group is generally larger than the control group, the higher the Cohen's d value, the larger the effect size.

\subsubsection{Dataset Effectiveness} \label{sec:exp_effective}

For \textbf{Q1}, we analyse the inference results of the overall variation of NetEaseCrowd presented in Table~\ref{tab: overall_performance}. The majority of baseline methods exhibit improved inferential performance in both accuracy and F1-score when compared to MV. This implies that these models have learned associations among workers, tasks, and annotations so that the inference processes have been facilitated and the estimations are more accurate. It demonstrates the effectiveness of NetEaseCrowd for benchmarking truth inference methods from a foundational perspective .

\subsubsection{Temporal Information}

The annotation abilities of workers are evolving over time. Considering that all previous techniques do not incorporate temporal information into their worker models and only use compromise values to represent worker abilities at each time step, we compare the inference differences between set-wise inference and overall inference to investigate \textbf{Q2}.

Set-wise inference refers to the inference procedure conducted on each task set. Note that each task set contains multiple tasks annotated by workers within a short time period, such as one hour or one day. A hypothesis is that workers' abilities could be relatively stable within a very short period, making compromise values more representative and accurate in representing worker ability. Meanwhile, more accurate ability values could lead to more precise inference results.

The comparison results, as displayed in Table~\ref{tab: overall_performance}, reveal that most methods exhibit improved performance when applied task set by task set rather than directly over the entire dataset. The Cohen's d values highlight a moderate difference caused by this improvement. This observation is consistent with the discrepancy distributions in Figure~\ref{fig: analysis-capwise-accuracy-diff}, indicating that relying on a single compromised value to represent worker accuracy across all timestamps is a suboptimal choice. The performance outcomes also suggest a more dependable compromise value for estimating workers' abilities when employed on a task-set basis. Overall, the above experimental results validate the possibility of leveraging the provided temporal information to better model worker ability and improve inference performance. It should be noted that in addition to task sets, we also provide the timestamp of each annotation, which would enable more fine-grained modeling in future studies.

\begin{figure}[t]
	\centering
	
	\begin{subfigure}{0.45\linewidth}
		\includegraphics[width=0.95\linewidth]{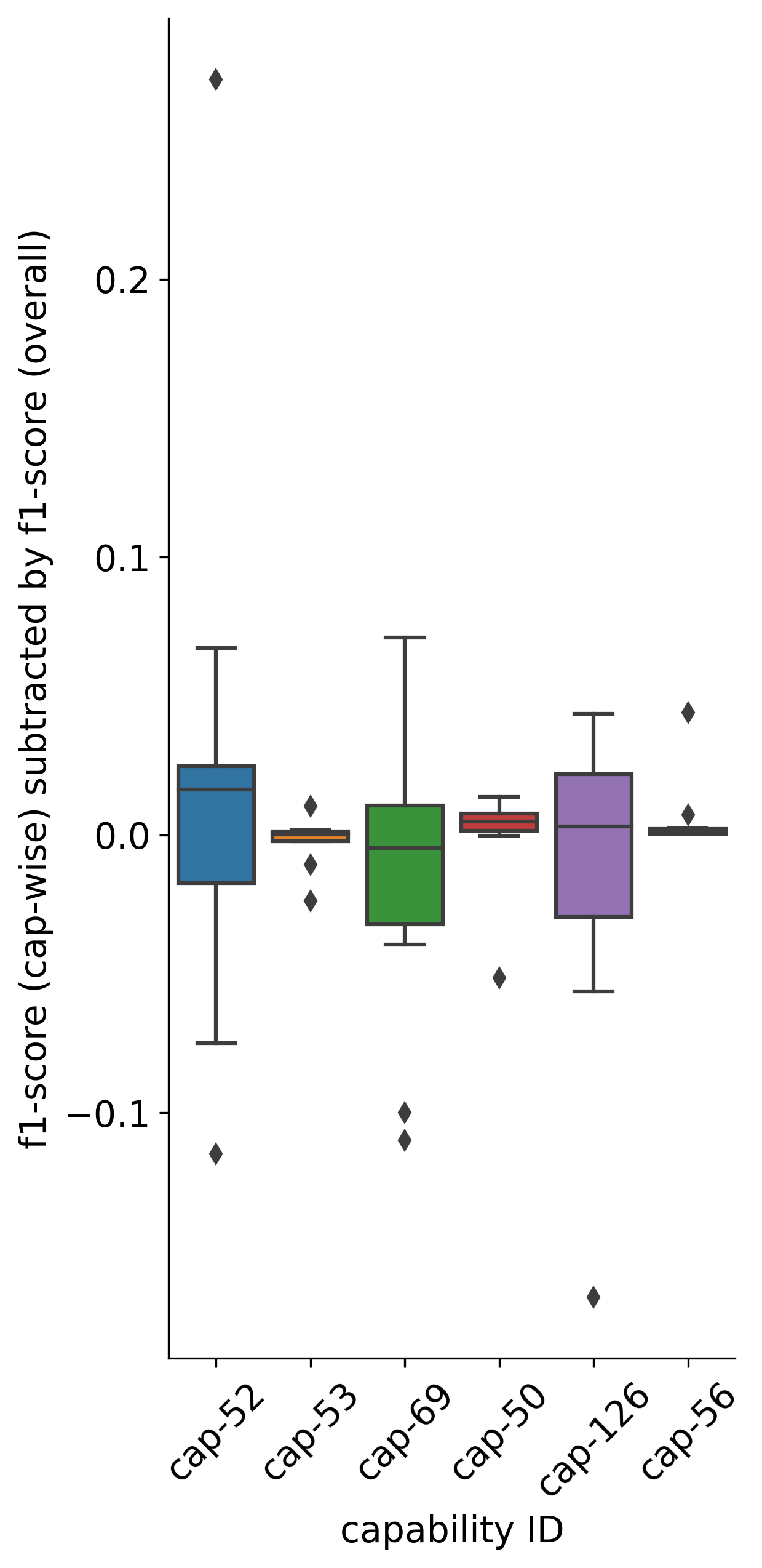}
		\caption{Differences in F1-score.}
		% \label{fig:redundancy_1}
	\end{subfigure}
	\begin{subfigure}{0.45\linewidth}
		\includegraphics[width=0.95\linewidth]{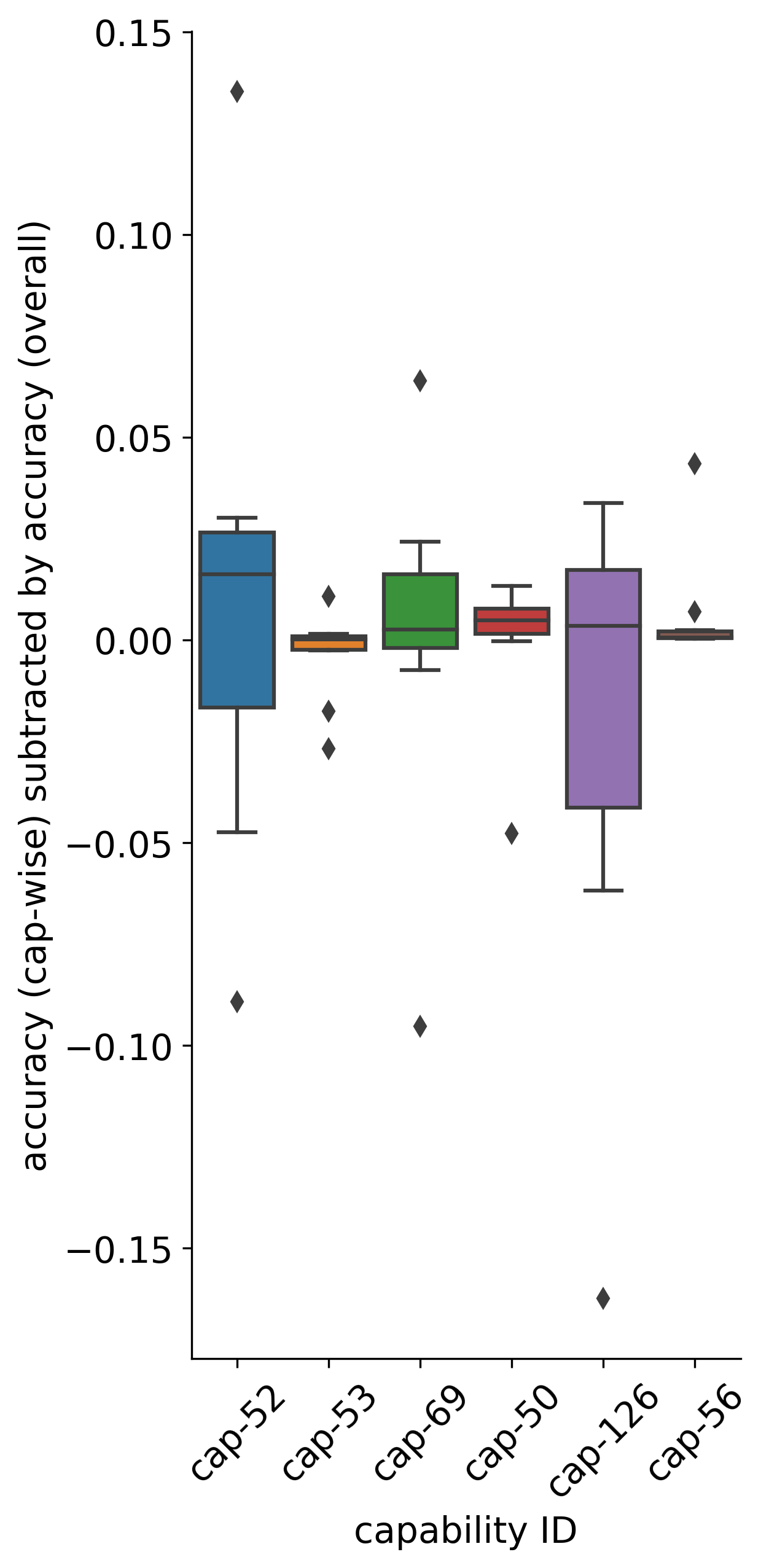}
		\caption{Differences in accuracy.}
		% \label{fig:redundancy_1}
	\end{subfigure}
	
	\caption{Performance comparisons between cap-wise and overall variations on the sub-datasets for each capability ID.}
	\label{fig: difference_distribution}
\end{figure}

\subsubsection{Capability Information} \label{sec:exp_cap}
Similar to the set-wise variation, in each cap-wise variation, the abilities of workers are possibly to be more stable rather than that in the overall variation, thus the performance of baseline methods is expected to be better in the cap-wise variation than in the overall variation. We hereby investigate \textbf{Q3} through the performance caparisons.

As shown in Table~\ref{tab: overall_performance}, the majority of baseline methods also demonstrate superior performance in cap-wise variation compared to overall variation. Cohen's d values suggest a small effect size from a statistical perspective, indicating that when inferring the labels of questions in NetEaseCrowd, employing previous methods task set by task set is more effective than applying them directly to the overall dataset. Considering the distribution differences of worker accuracy under different capability IDs in Figure~\ref{fig: cap-wise-accuracy-features}, it suggests that relying on a single compromised value to represent a worker's ability across different capabilities is also suboptimal. Thus, the performance results demonstrate the effectiveness of capability ID information in NetEaseCrowd.

Furthermore, we observe the performance differences of each method between cap-wise and overall variations under each capability ID. To be specific, after conducting inference based on each of these two variations, we compute inference metrics, accuracy and F1-score, for the tasks within every capability ID. Then, we subtract the performance metrics for cap-wise variation from the metrics for overall variation, whereby higher differences indicate better performance in the cap-wise variation. We show the distributions of these differences for different capability IDs in Figure~\ref{fig: difference_distribution}, the performance of cap-wise variation is either comparable to or surpasses that of overall variation across all capabilities.

\subsubsection{Supervised} \label{sec:exp_supervised}

\begin{table}[t]
    \centering
    \caption{Performance on \textbf{supervised} dataset. Duration in millisecond measures the inference time per single task.}
    \begin{tabular}{ccccc}
    \toprule
         Methods & Type & Accuracy& F1-score &  Duration (ms) \\\hline
         MV & w/o  & 0.93498 & 0.93502 & 0.00903 \\
        DS & w/  & 0.97623 & 0.97623 & 0.03819\\
        MACE & w/  & 0.97721 & 0.97721 & 1.64556\\
        Wawa & w/  & 0.96200 & 0.96200 & 0.02232\\
        ZeroBasedSkill & w/ & 0.96415 & 0.96414 & 0.53005 \\
        GLAD & w/  & 0.97416 & 0.97415 & 2.89311 \\
        EBCC & w/  & 0.97571 & 0.97571 & 0.07231\\
        ZC & w/  & 0.98074 & 0.98073 & 1.89806 \\
        TiReMGE & w/  & 0.93484 & 0.93488 & 4.87427\\
        LAA & w/ & 0.97353 & 0.97353 & 13.15747\\
        BiLA & w/ & 0.88416 & 0.88349 & 142.22800\\ 
        LAA-su & w/o & 0.83147 & 0.83154 & 0.00596 \\
        BiLA-su & w/o & 0.87884 & 0.87824 & 0.46704 \\
        
        \bottomrule
        
    \end{tabular}
    % \label{tab:taskwise_performance}
    \label{table:supervised-results}
\end{table}

For \textbf{Q4}, we investigate the performance of all methods in the supervised dataset, the results are presented in Table~\ref{table:supervised-results}. 
The \textit{Type} column indicates the inference manner of the method: `w/' denotes that during the inference, the corresponding method requires training the model with the target annotations data by solving the designed optimization problem, while `w/o' means the inference can be directly conducted without any further training. Therefore, methods categorized as `w/o' type are expected to have superior time efficiency during inference.

Most previous methods overlook this time-critical scenario. The basic method MV inherently falls into the `w/o' category. All DS-like methods such as MACE, Wawa, ZeroBasedSkill, GLAD, EBCC, and ZC are designed to solve the likelihood maximization problem during inference, thus they also fall into the `w/' category. For the neural network methods, TiReMGE~\cite{wu2023crowdsourcing}, LAA, and BiLA operate under an unsupervised setting and require retraining the model on new annotations before producing aggregated labels, placing them in the `w/' category as well. However, for LAA and BiLA, it is possible to utilize them solely for inference once they are trained, resulting in two variations, LAA-su and BiLA-su.

The results in Table~\ref{table:supervised-results} demonstrate that LAA-su achieves the fastest inference duration for per single task. It's worth noting that, as neural network methods, both LAA-su and BiLA-su can be executed in parallel on a massive scale, resulting in constant time complexity regardless of the size of the online tasks. Conversely, the time complexity of all other methods, except MV, is influenced by the scale of the target annotations. This indicates the potential of neural network-based methods to accelerate inference duration and facilitate real-world deployment. However, both LAA-su and BiLA-su experience significant performance degradation compared to their `w/' versions, indicating that without a specific design for this supervised setting, previous methods currently lack generalizability. This issue may stem from the overfitting during the training phase of BiLA and LAA, which primarily rely on maximum likelihood estimation. This emphasizes the necessity of further research on truth inference methods suitable for real-world online scenarios.

Regarding the dataset itself, it is large enough and includes temporal information of workers' annotations and the groundtruth of each task, which provides a more effective foundation for the design, training, and evaluation of supervised algorithms.

%% file: conclusion.tex
\section{Discussion and Conclusion}
In this work, we introduced a novel dataset for long-term and online crowdsourcing truth inference, NetEaseCrowd, which is collected from a commercial data crowdsourcing platform. Compared to existing public datasets, the advantage of NetEaseCrowd is threefold. \textbf{(1) Large time duration.} The data is collected over about six months, and the timestamps of each annotation are reserved. The analysis in \ref{sec:worker_property} implies that workers will keep labeling on the platform for a long time and new workers keep coming. With statistical analysis and experiments, we prove that workers' labeling abilities vary over time, which indicates that modeling a worker's ability as static is not adequate for long-term applications such as truth inference on a crowdsourcing platform. \textbf{(2) Multiple task types.} The crowdsourcing platform releases tasks of different types requiring different capabilities of workers. The statistical analysis in \ref{sec:worker_property} and the experimental analysis in \ref{sec:exp_cap} demonstrate the difference that workers perform in tasks related to different capabilities. \textbf{(3) Large data volume.} The data size is much larger than all the existing datasets and reflects the fact that a crowdsourcing platform could bear a heavy online computational burden. NetEaseCrowd is large enough to evaluate the efficiency of truth inference algorithms.

We hope this dataset will help with the development of truth inference algorithms as well as relevant applications such as task assignment. According to our analysis of the dataset, future studies include evaluating fine-grained worker abilities (relevant to different types of tasks), modeling the evolution of worker abilities over time, and highly efficient online algorithms. Considering the availability of a large amount of supervised data and the efficiency of the inference stage, half-supervised or supervised algorithms for truth inference might be a promising research direction, although there are still many challenges to address, as \ref{sec:exp_supervised} indicates.